# Microlens-Integrated Grating Couplers for Efficient Free-Space Coupling in Silicon Photonics

SIMON A. BONGARZ,[1,2,*] RYOSUKE MASHIKO,[2] ANNE R. KROO,[2] JAEKWON LEE,[2] CARSON G. VALDEZ,[2] DAVID A. B. MILLER,[2] AND OLAV SOLGAARD[2]

[1]*Department of Applied Physics, Stanford University, Stanford, California 94305, USA*
[2]*Department of Electrical Engineering, Stanford University, Stanford, California 94305, USA*
[*]*simon.bongarz@stanford.edu*

**Abstract:** Grating-coupler arrays are used to couple free-space light into photonic chips, but the space required for tapers and routing waveguides limits their optical fill factor. Microlens arrays can increase the effective fill factor by collecting light over a larger area and focusing it onto each grating. Separately fabricated lens arrays, however, require assembly and precise alignment to the chip. Here we integrate refractive microlenses directly with high-efficiency, normal-incidence grating couplers using two-photon polymerization. Each lens has a $100 \times 100$ µm$^2$ aperture and focuses incident light onto the grating. Writing the lens body at coarse resolution and the optical surface at fine resolution reduces the print time from 9 h to 19 min per lens. Under illumination by a Gaussian beam with a 276 µm mode-field diameter, a printed lens increases the measured coupled power by a factor of 33 (15.2 dB) relative to the same grating before printing. Loss normalization gives an estimated lens-to-mode efficiency of 68%. The lenses show no measurable change in coupling after three 3 h thermal cycles at 150 °C. These results support direct printing as an approach to increasing free-space collection while simplifying lens integration in photonic systems.

## 1. Introduction

Efficient coupling between free-space beams and on-chip waveguides is needed in a growing range of photonic integrated circuit (PIC) applications. These include light detection and ranging (LiDAR) [1, 2], free-space optical communication [3, 4], optical interconnects [5], quantum optics [6], chip-scale atomic systems [7, 8], and astrophotonics [9–11]. In each case, the interface must transfer light from a free-space field into the sub-micrometer mode of a single-mode waveguide. Loss at this interface can consume a substantial part of the system's optical budget.

Collection efficiency is especially important in photon-starved astronomical instruments. AstroPIC, a photonic integrated coronagraph under development for NASA's Habitable Worlds Observatory, must sample a telescope pupil and couple the sampled field into tens to hundreds of single-mode waveguides. Lower throughput increases the exposure time needed to reach a given signal-to-noise ratio and can reduce the expected exo-Earth yield [9, 12, 13].

Surface grating couplers provide convenient input ports: they can be placed across a wafer, tested before dicing, and arranged in two-dimensional arrays [14]. Normal incidence is useful for coupling to free-space beams, microlens arrays, multicore fibers, and vertical-cavity surface-emitting lasers. Conventional gratings, however, exhibit second-order back reflection and poor directionality at normal incidence [15]. Three-wave interaction gratings (TWIGs; Fig. 1a) address these losses using three scattering sites per local period. Their interference suppresses both back-reflected and substrate-directed waves without requiring additional material layers. Foundry-fabricated TWIG couplers on a standard 220 nm silicon-on-insulator platform have demonstrated a peak TE-mode coupling efficiency of 85.4% (−0.69 dB) at normal incidence [16, 17].

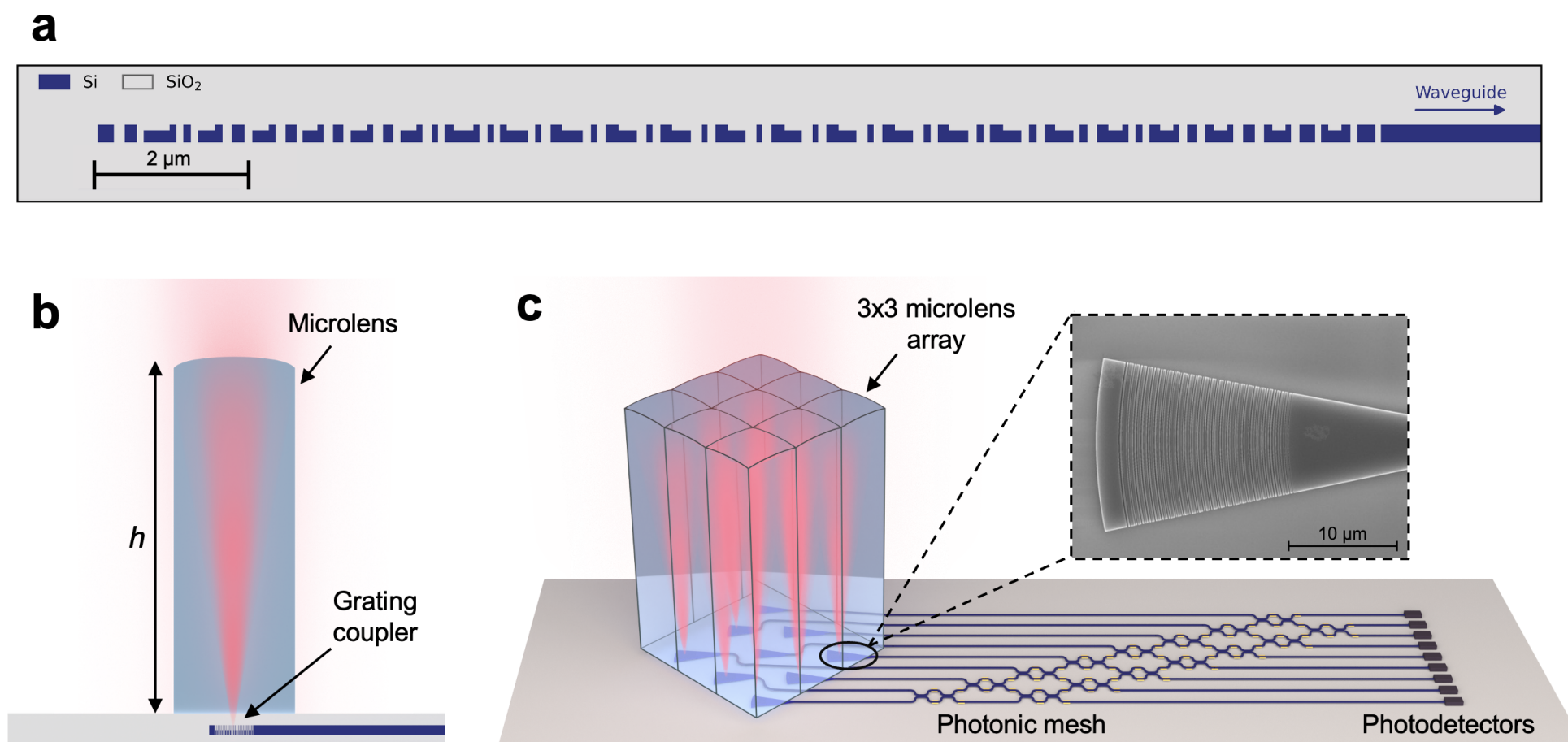

Fig. 1. **Concept for free-space coupling with integrated microlenses.** **(a)** Cross section of a dual-etch TWIG coupler designed for normal incidence. **(b)** Side view of a printed microlens of height *h*, focusing normally incident light onto the underlying grating. **(c)** Proposed array interface: a $3 \times 3$ microlens array focuses light onto nine grating couplers connected to a $9 \times 2$ photonic mesh with photodetector readout. Each lens is centered on its coupler. Inset: SEM image of a fabricated TWIG coupler.

A grating coupler operating at 1550 nm typically accepts a Gaussian mode with a diameter of about 10 µm, whereas practical arrays use pitches of tens to hundreds of micrometers to accommodate tapers and routing waveguides [14]. The active grating area therefore occupies only a small fraction of the array footprint. This low optical fill factor limits the power collected from a large beam, even when each grating couples its matched mode efficiently.

Microlenses address this mismatch by collecting light over a larger area and directing it to each input port. For example, the proposed AstroPIC architecture includes a microlens array to sample an image of the telescope pupil and couple light into the chip [9]. The focal-plane switch-array LiDAR in Ref. [1] places lenslets between its grating antennas and imaging optics. Separately fabricated lens arrays must be assembled and actively aligned to the chip. The resulting interface introduces alignment errors, Fresnel losses, and additional packaging steps.

Directly printing the optics on the chip incorporates lens alignment into fabrication. 3D printing can produce optical elements with nanometer-scale surface roughness [18]. Printed micro-optics have coupled dissimilar photonic components with efficiencies up to 88% and alignment tolerances compatible with passive assembly [19, 20]; the process is also used commercially for photonic packaging [21, 22]. Two-photon grayscale lithography has demonstrated surface roughness below 5 nm and automated registration to on-chip features with an accuracy of ±100 nm [23, 24]. Printed mirror–lens couplers at chip facets have redirected edge-coupled light upward for vertical fiber coupling, with a reported coupling loss of 1.2 dB [25].

Much of this work concerns fiber-to-chip packaging, including increased alignment tolerance and longer working distances for tilted grating couplers [26–28]. For free-space coupling, Zerboni *et al.* printed refractive microlenses above foundry gratings that radiate at 12.5° from the chip normal [29]. Their tilted circular lenses, positioned 30 µm above the gratings, reduced the emitted beam divergence from 100 mrad to below 20 mrad and provided a 15 dB coupling improvement for free-space beams with waists of several hundred micrometers.

Here we combine directly printed refractive microlenses with high-efficiency TWIG couplers to collect light at normal incidence. For a Gaussian input beam with a 276 µm mode-field diameter, we measure a 33-fold (15.2 dB) increase in coupled power on the same grating before and after lens printing. A two-tier writing process reduces fabrication time to 19 min per lens on a Nanoscribe Photonic Professional GT, and the printed lenses retain their optical performance after three 3 h cycles at 150 °C. We also fabricate $3 \times 3$ lens arrays and measure their surface profiles. Optical coupling is evaluated on individual lenses; demonstrating array coupling requires a chip with a matching grating array.

## 2. Device concept

The interface combines a normal-incidence TWIG coupler with a solid polymer microlens printed directly above it (Fig. 1a,b). The coupler accepts a Gaussian mode with a 10.4 µm mode-field diameter (MFD) and has demonstrated sub-decibel insertion loss [16, 17]. The lens collects light across a larger aperture. Its aspheric upper surface refracts the light at the air–polymer interface and focuses it through the polymer onto the grating. We design the focal field to match the mode accepted by the coupler.

Each lens is aligned to its grating during printing, eliminating the separate lens-array assembly and active lens-to-chip alignment steps [19, 21]. The lens expands the beam emerging from the small grating coupler, relaxing lateral positioning tolerances for external optics at the expense of tighter angular alignment. This trade-off benefits packaging e.g. for optical interconnects, since angular alignment is generally easier to control mechanically than micrometer-scale positioning. An array of these interfaces could sample an extended beam and feed a photonic mesh with integrated photodetectors, as illustrated in Fig. 1c. The square $100 \times 100$ µm$^2$ aperture allows neighboring lenses to tile at a 100 µm pitch while providing space beneath each lens for the coupler, taper, and routing waveguides.

The focal distance must be chosen to match the diffraction pattern of this aperture to the grating mode. For a uniformly illuminated square aperture of side $D$, focused at a distance $f$ in a medium of refractive index $n$, the focal field is [30]

$$E(x, y) \propto \operatorname{sinc}\left(\frac{\pi D x}{\lambda_n f}\right) \operatorname{sinc}\left(\frac{\pi D y}{\lambda_n f}\right), \qquad \lambda_n = \frac{\lambda}{n}. \tag{1}$$

Here, $\lambda$ is the free-space wavelength. The maximum overlap of this field with a Gaussian mode of waist $w_0$ is approximately 79%. For comparison, the corresponding limit for the Airy pattern of a circular aperture is approximately 81% [31]. The square aperture reaches its overlap limit when

$$\frac{\pi D w_0}{2\lambda_n f} \approx 1 \quad \Longleftrightarrow \quad f_{\text{opt}} \approx \frac{\pi n D w_0}{2\lambda}. \tag{2}$$

For $D = 100$ µm, $w_0 = 5.2$ µm, $\lambda = 1550$ nm, and $n = 1.48$ for IP-S [32], Eq. (2) gives $f_{\text{opt}} \approx 780$ µm. The numerical optimization described in Section 3 gives a focal length of 750 µm. Because the focus lies at the grating plane at the base of the lens, the focal length also sets the lens height.

A larger aperture collects more power per channel but requires a taller lens and a longer print time. It also reduces the number of channels that fit across a beam of fixed size. We chose a 100 µm aperture to balance collection area, fabrication time, and the space required for waveguide routing.

## 3. Microlens design and fabrication

We optimized the refracting surface in Ansys Zemax OpticStudio using an even asphere. The conic and higher-order terms correct spherical aberration, and the freeform capability of two-photon polymerization (2PP) allows the resulting surface to be printed directly. Its profile is

$$z(r) \;=\; \frac{cr^2}{1+\sqrt{1-(1+k)\,c^2r^2}} \;+\; \sum_{i=1}^{5} \alpha_{2i}\, r^{2i}, \tag{3}$$

where $z$ is the surface sag relative to the vertex, $r$ is the radial distance from the optical axis, $c = 1/R$ is the vertex curvature, $R$ is the vertex radius of curvature, and $k$ is the conic constant. The polynomial coefficients $\alpha_{2i}$ extend through tenth order in $r$.

The model uses the refractive index of cross-linked IP-S at the design wavelength ($n = 1.48$ at $\lambda = 1550$ nm [32]). A flat-top input beam overfills the lens, producing uniform illumination within its square aperture. At the grating plane, the target field is a Gaussian mode with a 10.4 µm MFD and no wavefront curvature. We used the physical optics propagation (POP) module and its POP data operand to calculate the overlap of the propagated field with this target mode, then maximized the overlap. Optimization took approximately 15 min.

The optimized lens has a $100 \times 100$ µm$^2$ aperture, a height of 750 µm, and a nominal vertex radius of curvature of 251 µm. Its simulated lens-to-mode efficiency is 79%, consistent with the square-aperture overlap limit in Section 2. Figure 2a shows the aspheric surface, and Fig. 2b shows the resulting irradiance at the grating plane.

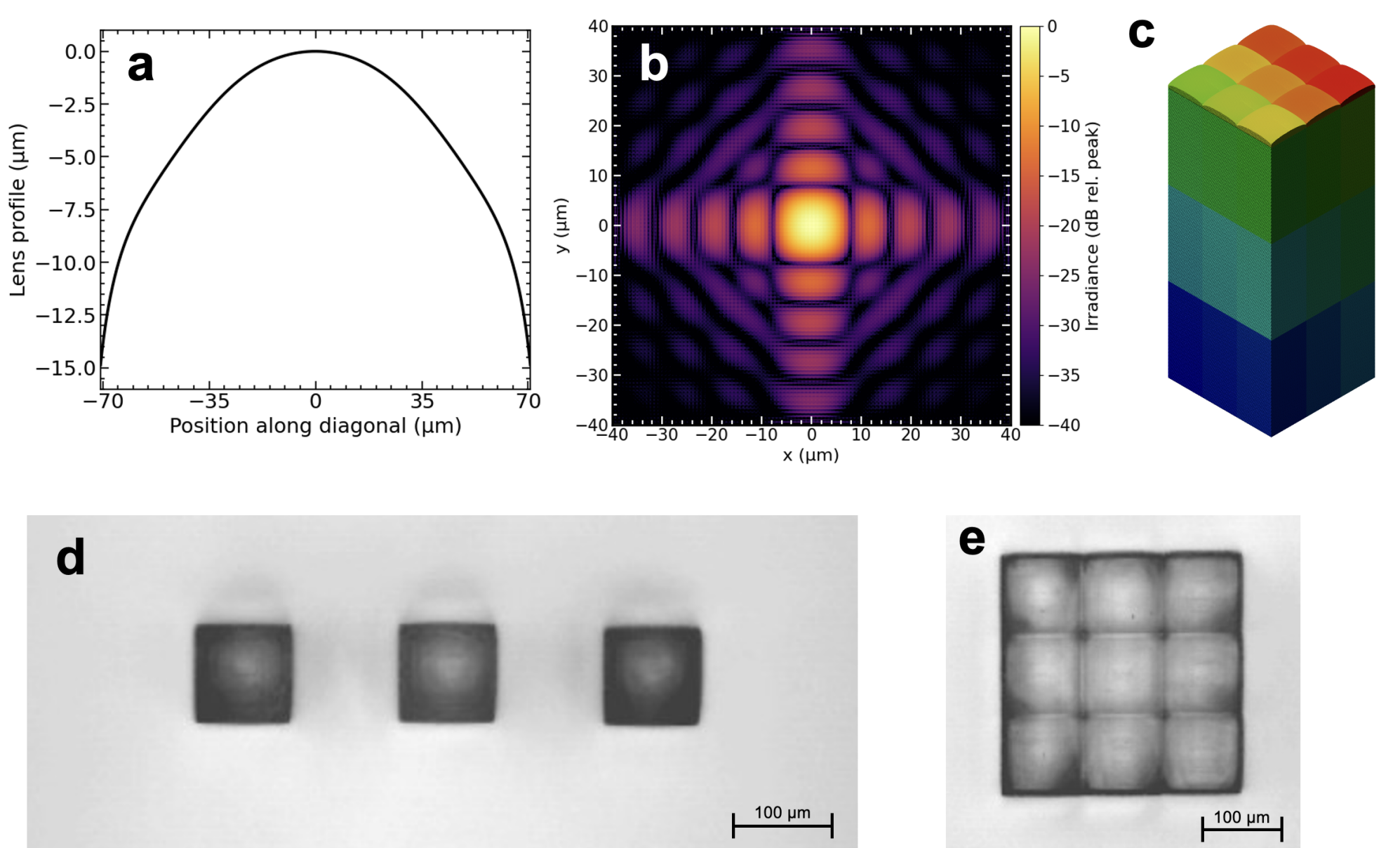

Fig. 2. **Microlens design and fabrication.** **(a)** Optimized surface profile along the lens diagonal, with sag measured relative to the vertex. **(b)** Simulated irradiance at the grating plane for uniform illumination of the square aperture, plotted in dB relative to the peak. **(c)** Two-tier print job for a $3 \times 3$ lens array. The lower 730 µm of each lens is written at coarse resolution and the upper 20 µm at fine resolution. Color indicates progress through the 171-min print, from blue at the start to red at the end. **(d), (e)** Microscope images of printed $3 \times 1$ and $3 \times 3$ lens arrays, respectively.

We fabricated the lenses on the photonic chip characterized in Ref. [17], which contains a linear array of grating couplers. After placing a drop of IP-S resist (Nanoscribe GmbH) on the chip, we printed the lenses with a Nanoscribe Photonic Professional GT in dip-in laser lithography mode, using a 25×, NA-0.8 objective. The integrated camera and on-chip markers provided lens-to-grating alignment with approximately ±1 µm accuracy, limited by camera resolution and marker size.

Most of the lens height serves as a spacer between the refracting surface and the grating. We therefore divided the print into two tiers: the lower 730 µm used 0.5 µm slicing and hatching, whereas the upper 20 µm used 0.1 µm slicing and hatching. The fine tier covers the full optical surface, whose sag is approximately 15 µm (Fig. 2a). Concentrating the fine writing in this region reduced the print time from 9 h to 19 min per lens, or 171 min for a $3 \times 3$ array (Fig. 2c).

We slightly reduced the laser power in the fine tier to maintain a uniform exposure dose. Power at the chip surface also required care: early prints with the full laser power focused on silicon damaged the gratings through absorption and heating. Lowering the power and moving the focus slightly above the grating reduced the intensity at the silicon in subsequent prints.

After printing, we removed unexposed resist by developing the structures for 15 min in PGMEA-based SU-8 developer. We then rinsed them for 15 min in isopropyl alcohol, dried them with nitrogen, and completed polymer cross-linking with a 30 s ultraviolet flood exposure at approximately 365 nm. Figures 2d,e show the finished $3 \times 1$ and $3 \times 3$ arrays.

The reported print times were obtained on the Photonic Professional GT, an early commercial 2PP platform. Newer two-photon grayscale systems report throughput gains of up to 60-fold over conventional 2PP [33]. Based on those gains, we estimate that a Nanoscribe Quantum X could print each lens in less than 1 min, placing arrays of hundreds of lenses within a few hours of printing.

Polymerization and development can shrink printed structures, while voxel size, slicing, hatching, and exposure dose can introduce repeatable shape errors [34]. To evaluate the effect of fabrication errors on coupling efficiency, we varied the lens height and vertex radius of curvature independently by ±20 µm in the model (Fig. 3a). The nominal values were 750 µm and 251 µm, respectively. A ±20 µm change in height altered efficiency by less than 1 percentage point. A ±5 µm change in radius reduced efficiency by about 3 percentage points. Within the simulated range, efficiency fell below 60% only for radii approximately 15–20 µm smaller than the nominal value.

Lateral alignment requires tighter control (Fig. 3b). Offsets of 500 nm along both $x$ and $y$ reduced the simulated efficiency from 79% to 78%. For offsets of 1 µm along each axis, representative of our alignment accuracy, efficiency was approximately 73%; offsets of 3 µm along each axis reduced it to 40%. The ±100 nm registration accuracy reported for newer Nanoscribe systems [23, 24] would correspond to an alignment loss below 0.1 dB.

We measured the printed profiles with a Zygo NewView 9000 coherence-scanning white-light interferometer. A 50× objective provided a 170 µm field of view. We stitched adjacent fields to cover a $3 \times 3$ array, corrected the height maps for tilt, and compared the measured surfaces with the nominal design.

With a numerical aperture of approximately 0.55, the objective cannot reliably measure surface slopes greater than about 33°. The slopes near the lens corners reach approximately 55°, exceeding this limit. The map in Fig. 3c therefore lacks data at the corners and covers about 12 µm of the full 15 µm sag. The $x$ and $y$ centerline profiles in Fig. 3d pass through regions with lower slopes.

We used these measurements to correct the printed shape in two steps. First, scaling all design dimensions up by 3% compensated for polymer shrinkage. The resulting $3 \times 3$ array still showed a mean deviation of 1.2 µm at a radius of 45 µm, near the aperture edge. We then subtracted the measured deviation from the design before slicing, following a profile precompensation approach [34]. Reprinting and measuring under the same conditions reduced the mean deviation at that radius to 170 nm, a sevenfold improvement. Figures 3c,d show the compensated array and its residual profile errors.

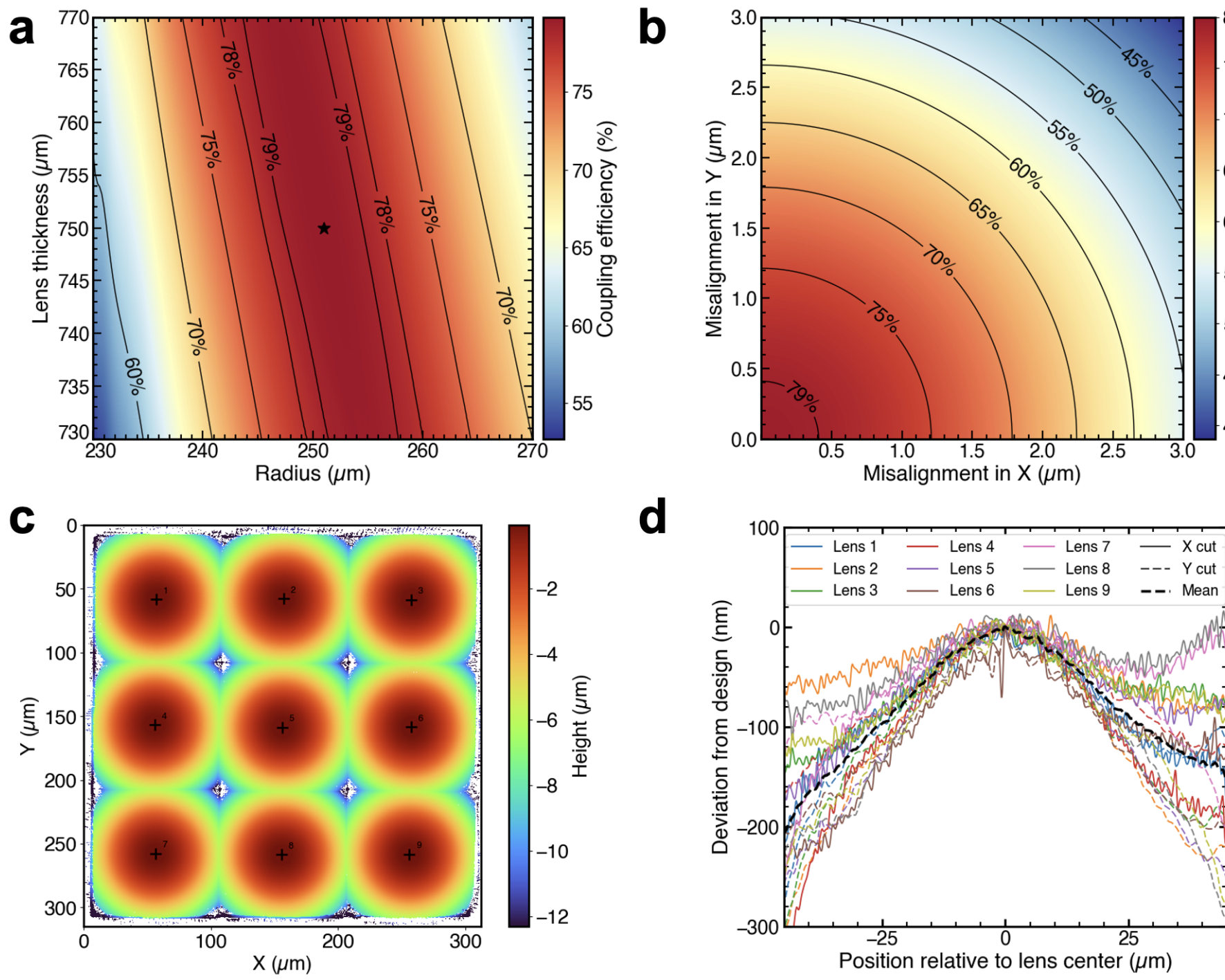

Fig. 3. **Fabrication tolerances and surface measurements.** **(a)** Simulated coupling efficiency as the lens height and vertex radius of curvature vary by ±20 µm about their design values. The black star marks the nominal height of 750 µm and radius of 251 µm. **(b)** Simulated efficiency for lateral lens-to-grating offsets of up to 3 µm in $x$ and $y$. **(c)** Tilt-corrected height map of the compensated $3 \times 3$ array measured by optical profilometry. Crosses mark the lens centers. **(d)** Residual deviations from the design along the $x$ (solid) and $y$ (dashed) centerlines of all nine lenses. The black dashed curve shows the mean deviation.

Simulations incorporating the measured residual error predict a coupling efficiency slightly below 70% under uniform illumination, compared with 79% for the ideal surface. Thus, the remaining shape error can appreciably reduce mode overlap even after compensation. Further correction iterations or improved printing resolution may bring the fabricated surface closer to the design.

## 4. Experimental results

We measured optical coupling and thermal stability on individual microlenses. Figure 4a shows the free-space setup, which illuminates the chip at normal incidence. A collimated short-wave infrared (SWIR) LED and relay optics provide a magnified image of the chip on an InGaAs camera, resolving individual lenses and gratings. A separate tunable laser supplies the beam for coupling measurements. Lenses in the laser path set the beam size, and rotation of its polarization-maintaining fiber aligns the polarization with the grating's TE mode.

Light entering the input grating travels through a single-mode waveguide to a second grating, where an output fiber carries it to a photodetector. We used this same downstream waveguide and collection path for the measurements before and after lens printing. The detected-power ratio therefore provides a measure of the improvement at the free-space input interface.

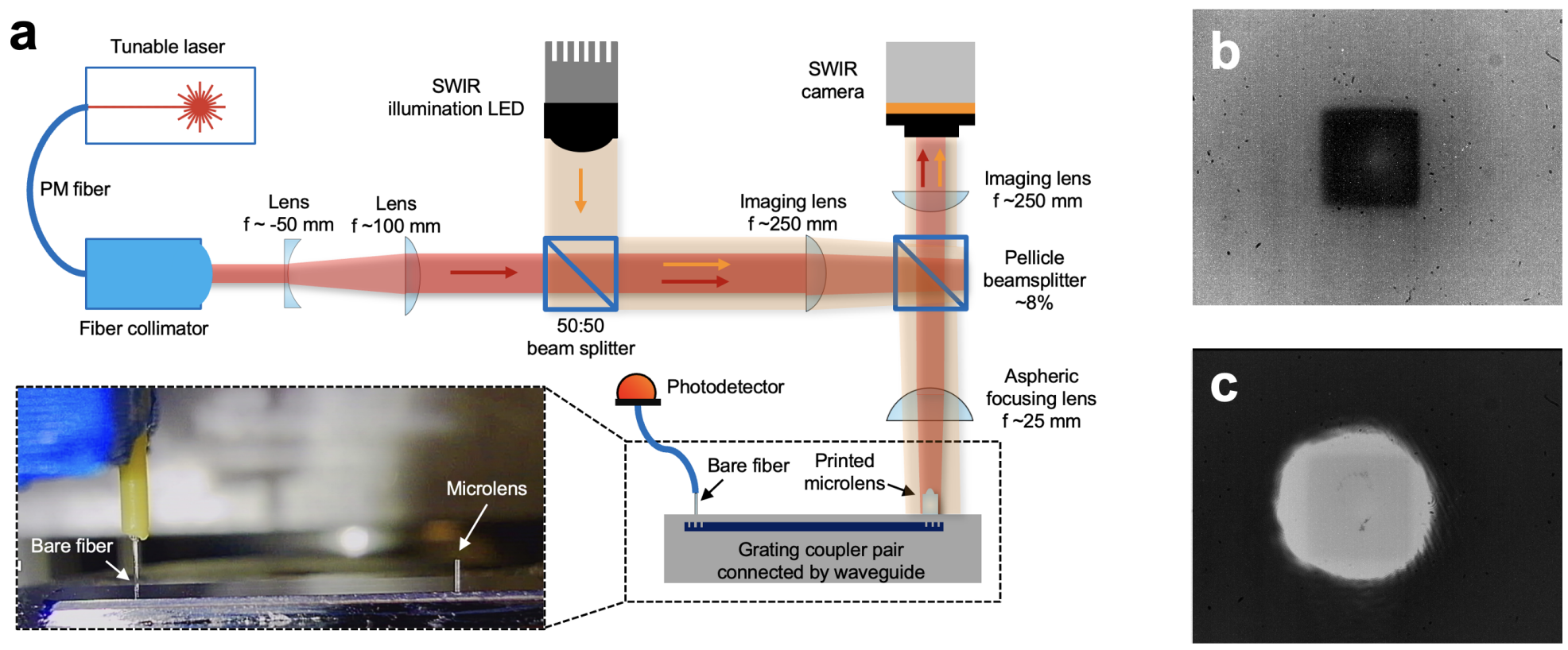

Fig. 4. **Setup for free-space coupling measurements.** **(a)** Optical paths for SWIR imaging, LED illumination, and tunable-laser excitation of the chip. An output fiber collects the transmitted light. Inset: photograph of the output fiber and printed microlens. **(b)** SWIR image of a printed lens under LED illumination. **(c)** Composite image combining an LED-illuminated view of the lens with the beam emerging under back illumination through the output fiber.

Alignment proceeded in three steps. We first positioned the chip on a rotation stage and located the lens under LED illumination (Fig. 4b). We then injected laser light through the output fiber and scanned its position until the beam emerging from the lens was visible on the camera (Fig. 4c). Finally, we switched the laser to the free-space input path and adjusted the beam position over the lens. This alignment procedure took only a few minutes. We jointly optimized the input-beam and output-fiber positions for maximum detected power before recording spectra from 1500 to 1580 nm in 0.1 nm steps.

Both device configurations were measured under the same illumination conditions. The incident beam had an approximately Gaussian profile with a 276 µm MFD (Fig. 5b) and a total power of 70 µW, measured immediately before the chip. Figures 5a,c show the bare and microlensed couplers under this illumination.

At 1550 nm, the detected peak power increased from 116 nW before lens printing to 3.85 µW afterward, a 33-fold enhancement (15.2 dB). Measuring the same grating in both configurations removes variation between fabricated couplers from this comparison. Figure 5d shows the corresponding spectra.

To interpret the enhancement, we compared the measured powers with a loss model for each configuration. For the bare grating, only 392 nW of the 70 µW beam falls within the grating aperture. Accounting for the measured grating efficiency, waveguide propagation loss [17], and other system losses predicts 137 nW at the detector. This is 35% of the power incident within the grating aperture. The measured value of 116 nW is lower, consistent with additional loss from residual angular alignment errors at the input beam and output fiber, and lateral fiber-to-grating misalignment.

The power incident on the lens aperture is 19.8 µW, approximately 50 times that incident on the bare grating aperture. Using the simulated lens-to-mode efficiency of 82% for the measured Gaussian beam, the Fresnel loss at the air–polymer interface, and the same downstream losses gives a predicted detected power of 5.47 µW, compared with a measured value of 3.85 µW.

We estimated the fabricated lens efficiency by normalizing the lensed measurement to the bare-grating result. The bare grating delivers 116 nW compared with a predicted 137 nW;

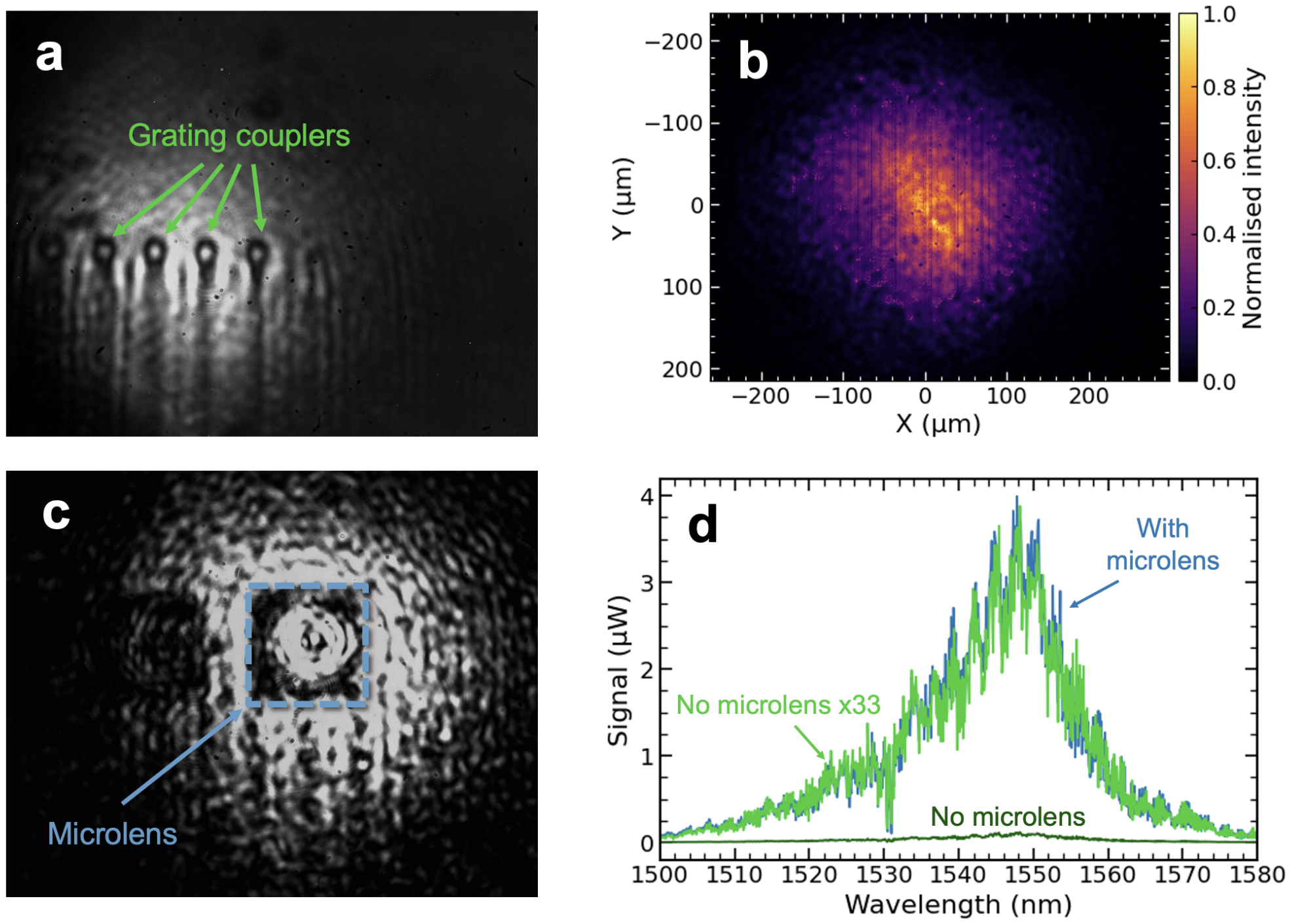

Fig. 5. **Measured enhancement in free-space coupling.** **(a)** Bare grating couplers under laser illumination. **(b)** Incident-beam profile measured with an InGaAs camera, approximated by a Gaussian with a 276 µm MFD. **(c)** The microlensed grating under the same illumination. **(d)** Throughput spectra measured before and after lens printing with a 70 µW incident beam. Multiplying the bare-grating spectrum by 33 gives close agreement with the microlensed spectrum.

applying this measured-to-predicted ratio to the lensed loss model gives a lens-to-mode efficiency of approximately 68%.

The ideal efficiency also depends on the incident intensity profile. The 79% design value applies to uniform illumination. For the measured Gaussian beam, whose intensity decreases toward the aperture edges, the simulated efficiency of an ideal lens is 82%, compared with an inferred efficiency of 68% for the fabricated lens. The simulations in Section 3 indicate that residual surface errors and lateral misalignment between the lens and the grating coupler contribute to this difference.

The enhancement remains approximately 33-fold over the measured 1500–1580 nm band. Scaling the bare-grating spectrum in Fig. 5d by 33 closely reproduces the microlensed spectrum, indicating no measurable additional spectral dependence from the lens over this range. The weak modulation present in both spectra arises from a Fabry–Perot cavity between the two grating couplers [17].

Subsequent packaging may expose printed lenses to heat during wire bonding of integrated modulators, heaters, or detectors, and during epoxy curing for fiber attachment or die mounting. To assess stability under such conditions, we placed a chip with printed lenses on a wire-bonding stage (Fig. 6a) and applied three thermal cycles. Each cycle comprised a ramp of 1 °C/s to 150 °C, a 3 h hold, and passive cooling in air.

The 150 °C bake matches typical stage temperatures for thermosonic Au wire bonding [35, 36] and epoxy curing [37, 38], and the cumulative 9 h hold is far longer than the time spent at temperature in either process. This temperature remains well below the reported onset of thermal yellowing in IP-S at approximately 250 °C, where increasing bulk absorption reduces transparency [39]. After cycling, we observed no visible change in the lenses and no change in coupling within measurement uncertainty (Fig. 6b).

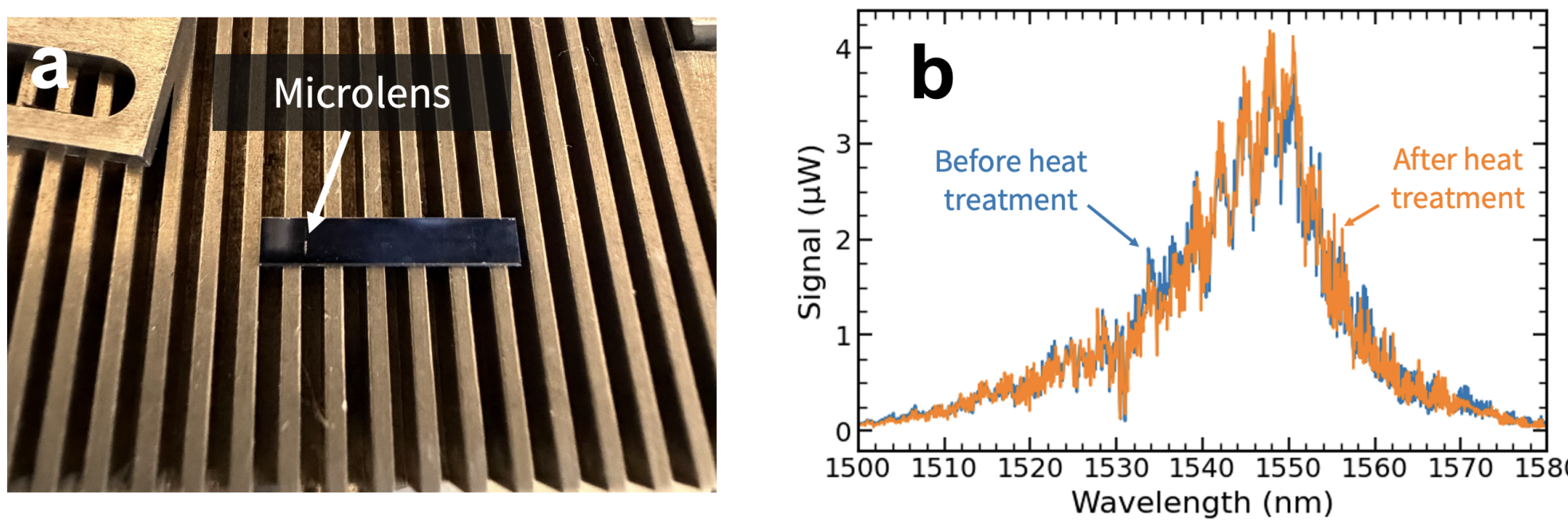

Fig. 6. **Optical performance before and after thermal cycling.** **(a)** Chip on the wire-bonding stage during three cycles of 3 h at 150 °C, chosen to assess exposure relevant to wire bonding and epoxy curing. **(b)** Throughput spectra of the microlensed coupler before and after cycling, showing no measurable change.

## 5. Conclusion

Directly printed refractive microlenses increase the collection area of high-efficiency, normal-incidence TWIG couplers. For a Gaussian input beam with a 276 µm MFD, we measured a 33-fold (15.2 dB) increase in coupled power on the same grating before and after printing. Loss normalization gives an estimated lens-to-mode efficiency of 68%, compared with a simulated ideal of 82% under the same illumination. Surface measurements and tolerance simulations identify residual profile error and lateral alignment as likely sources of the remaining loss.

The fabrication process combines coarse writing in the lens body with fine writing at the optical surface, reducing print time from 9 h to 19 min per lens. Measured-profile precompensation reduced the mean surface deviation near the aperture edge from 1.2 µm to 170 nm. The lenses also retained their coupling performance after three 3 h cycles at 150 °C. Together, these results establish a practical basis for integrating collection optics directly on photonic chips, with lens placement incorporated into fabrication.

The next step is to evaluate this approach across an array of optical channels. We have fabricated and profiled $3 \times 3$ lens arrays, while the coupling measurements reported here concern individual lenses. A chip with TWIG coupler arrays is in fabrication. Measurements on matching $N \times N$ lens and coupler arrays will be needed to quantify aggregate collection efficiency, effective fill factor, and interchannel crosstalk, and to assess integration with multiport photonic meshes.

**Funding.** NASA Ames Research Center (80NSSC24M0033); J. Lee was supported by the Stanford Energy Postdoctoral Fellowship and the Precourt Institute for Energy. D.A.B.M. was supported by the Air Force Office of Scientific Research (FA9550-23-1-0307).

**Acknowledgment.** We thank Dan Sirbu, Kevin Fogarty, Ruslan Belikov, Rachel Morgan, and Eduardo Bendek-Selman of the Astrophysics Branch, NASA Ames Research Center, for many insightful discussions during the course of this work, and Andrea Melloni, Francesco Morichetti, and Lorenzo Zerboni of Politecnico di Milano for stimulating interactions and insights. Part of this work was performed at nano@stanford (RRID:SCR_026695).

**Disclosures.** The authors declare no conflicts of interest.

**Data availability.** Data underlying the results presented in this paper are not publicly available at this time but may be obtained from the authors upon reasonable request.